\documentclass{iaus}
\usepackage{graphicx}

\title[$^6$Li in metal-poor halo stars] %% give here short title %%
{$^6$Li in metal-poor halo stars: real or spurious?}

\author[M. Steffen et al.]   %% give here short author list %%
{M.~Steffen$^1$,
R.~Cayrel$^2$, P.~Bonifacio$^{2,3,4}$, H.-G.~Ludwig$^{2,3}$, E.~Caffau$^2$}

\affiliation{$^1$Astrophysikalisches Institut Potsdam, Potsdam, Germany \\[\affilskip] 
$^2$GEPI, Observatoire de Paris/Meudon, France \\[\affilskip]
$^3$CIFIST Marie Curie Excellence Team, Observatoire de Paris/Meudon, France \\[\affilskip]
$^4$INAF, Observatorio Astronomico di Trieste, Trieste, Italy}

\pubyear{2009}
\volume{265}  %% insert here IAU Symposium No.
\pagerange{119--126}
\jname{Chemical Abundances in the Universe: Connecting First Stars to Planets}
\editors{K. Cunha, M. Spite \& B. Barbuy, eds.}
\begin{document}

\maketitle

\begin{abstract}
The presence of convective motions in the atmospheres of metal-poor halo 
stars leads to systematic asymmetries of the emergent spectral line profiles.
Since such line asymmetries are very small, they can be safely ignored for
standard spectroscopic abundance analysis. However, when it comes
to the determination of the $^6$Li/$^7$Li isotopic ratio, 
$q$(Li)=$n$($^6$Li)/n($^7$Li), the intrinsic
asymmetry of the $^7$Li line must be taken into account, because its
signature is essentially indistinguishable from the presence of a weak 
$^6$Li blend in the red wing of the $^7$Li line. In this contribution we
quantity the error of the inferred $^6$Li/$^7$Li isotopic ratio that arises 
if the convective line asymmetry is ignored in the fitting of the 
$\lambda\,6707$~\AA\ lithium blend. Our conclusion is that $^6$Li/$^7$Li 
ratios derived by \cite[Asplund \etal\ (2006)]{A2006}, using symmetric line 
profiles, must be reduced by typically $\Delta q$(Li) $\approx 0.015$. 
This diminishes the number of certain $^6$Li detections from 9 to 4 stars
or less, casting some doubt on the existence of a $^6$Li plateau.
\end{abstract}

\firstsection % if your document starts with a section,
\section{Introduction}
The spectroscopic signature of the presence of $^6$Li in the
atmospheres of metal-poor halo stars is a subtle extra depression in
the red wing of the $^7$Li doublet, which can only be detected in
spectra of the highest quality. Based on high-resolution, high
signal-to-noise VLT/UVES spectra of 24 bright metal-poor stars,
\cite[Asplund \etal\ (2006)]{A2006} report the detection of $^6$Li 
in nine of these objects. The average $^6$Li/$^7$Li isotopic ratio in 
the nine stars in which $^6$Li has been detected is $q$(Li) $\approx 0.04$ 
and is very similar in each of these stars, defining a $^6$Li plateau at 
approximately $\log n(^6$Li$) = 0.85$ (on the scale $\log n($H$) = 12$).
A convincing theoretical explanation of this new $^6$Li plateau
turned out to be problematic. Even when the depletion of the $^6$Li
isotope during the pre-main-sequence phase would be ignored, the high
abundances of $^6$Li at the lowest metallicities cannot be explained
by current models of galactic cosmic-ray production (for a concise 
review see e.g.\ \cite[Christlieb 2008]{C2008}, and references therein).

A possible solution of the so-called `second Lithium problem' was 
suggested by \cite[Cayrel \etal\ (2007)]{C2007}, who point out that 
the intrinsic line asymmetry caused by convection in the photospheres 
of cool stars is almost indistinguishable from the asymmetry produced 
by a weak $^6$Li blend on a presumed symmetric $^7$Li profile. 
As a consequence, the derived $^6$Li abundance should be significantly
reduced when the intrinsic line asymmetry in properly taken into
account. Using 3D non-LTE line formation calculations based on 3D 
hydrodynamical model atmospheres computed with the CO$^5$BOLD code 
(\cite[Freytag \etal\ 2002]{F2002}, \cite[Wedemeyer \etal\ 
2004]{W2004}, see also 
{\tt http://www.astro.uu.se/$\sim$bf/co5bold\_main.html}), 
we quantify the theoretical effect of the convection-induced line 
asymmetry on the resulting $^6$Li abundance as a function of effective 
temperature, gravity, and metallicity, for a parameter range that covers 
the stars of the \cite[Asplund \etal\ (2006)]{A2006} sample.

\section{3D hydrodynamical simulations and spectrum synthesis}
The hydrodynamical atmospheres used in the present study are part of the
CIFIST 3D model atmosphere grid, as described by \cite{L2009}. They have
been obtained from realistic numerical simulations with the CO$^5$BOLD code 
which solves the time-dependent equations of compressible hydrodynamics in 
a constant gravity field together with the equations of non-local, 
frequency-dependent radiative transfer in a Cartesian box representative of
a volume located at the stellar surface. The computational domain is periodic
in $x$ and $y$ direction, has open top and bottom boundaries, and is resolved 
by typically 140$\times$140$\times$150 grid points. The vertical optical depth
of the box varies from $\log \tau_{\rm Ross} \approx -8$ (top) to 
$\log \tau_{\rm Ross} \approx +8$ (bottom). The selected models cover 
the stellar parameter range $5900$~K~$< T_{\rm eff} < 6500$~K, 
$3.5 < \log g < 4.5$, $-3.0 <$ [Fe/H] $< -1.0$.

Each of the selected models is represented by about 20 snapshots chosen from 
the full time sequence of the corresponding simulation. All these 
representative snapshots are processed by the non-LTE code NLTE3D that solves
the statistical equilibrium equations for a 17 level lithium atom with 34 
line transitions, fully taking into account the 3D thermal structure of
the respective model atmosphere. The photo-ionizing radiation field is 
computed at $704$ frequency points between $\lambda\,925$ and 32\,407~\AA,
using the opacity distribution functions of \cite{C2004} to allow for 
metallicity-dependent line-blanketing, including the
H\,I--H$^+$ and H\,I--H\,I quasi-molecular absorption near $\lambda\,1400$ 
and $1600$~\AA, respectively. Collisional ionization by neutral hydrogen via 
the charge transfer reaction 
H($1s$) + Li($n\ell$) $\leftrightarrow$ Li$^+$($1s^2$) + H$^-$ is treated 
according to \cite{barklem2003}. More details are given in \cite{S2009}.

\begin{figure}[t]
%\vspace*{-2.0 cm}
\begin{center}
\mbox{\includegraphics*[bb=44 38 560 340,width=0.48\textwidth]
{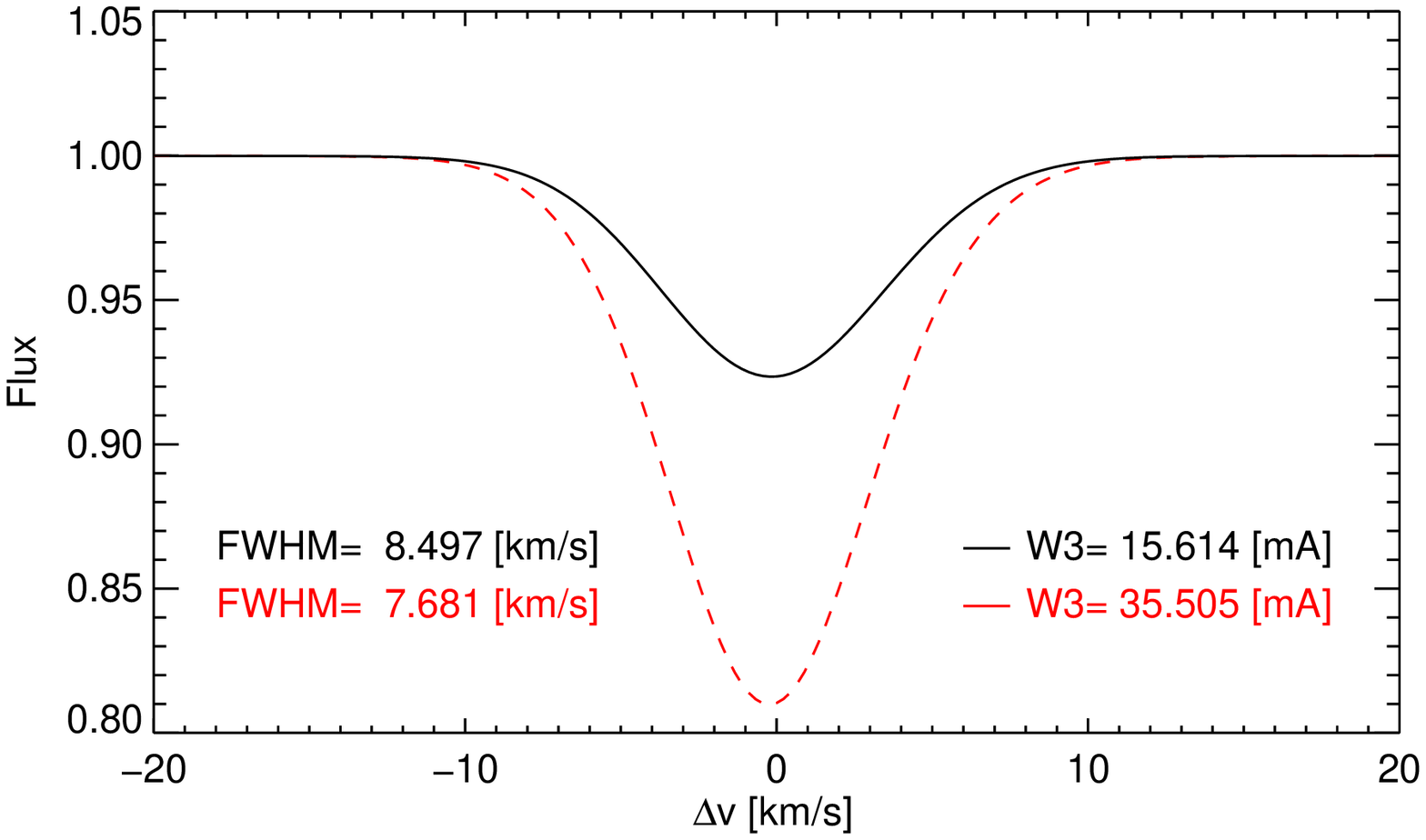}}
\mbox{\includegraphics*[bb=44 38 560 340,width=0.48\textwidth]
{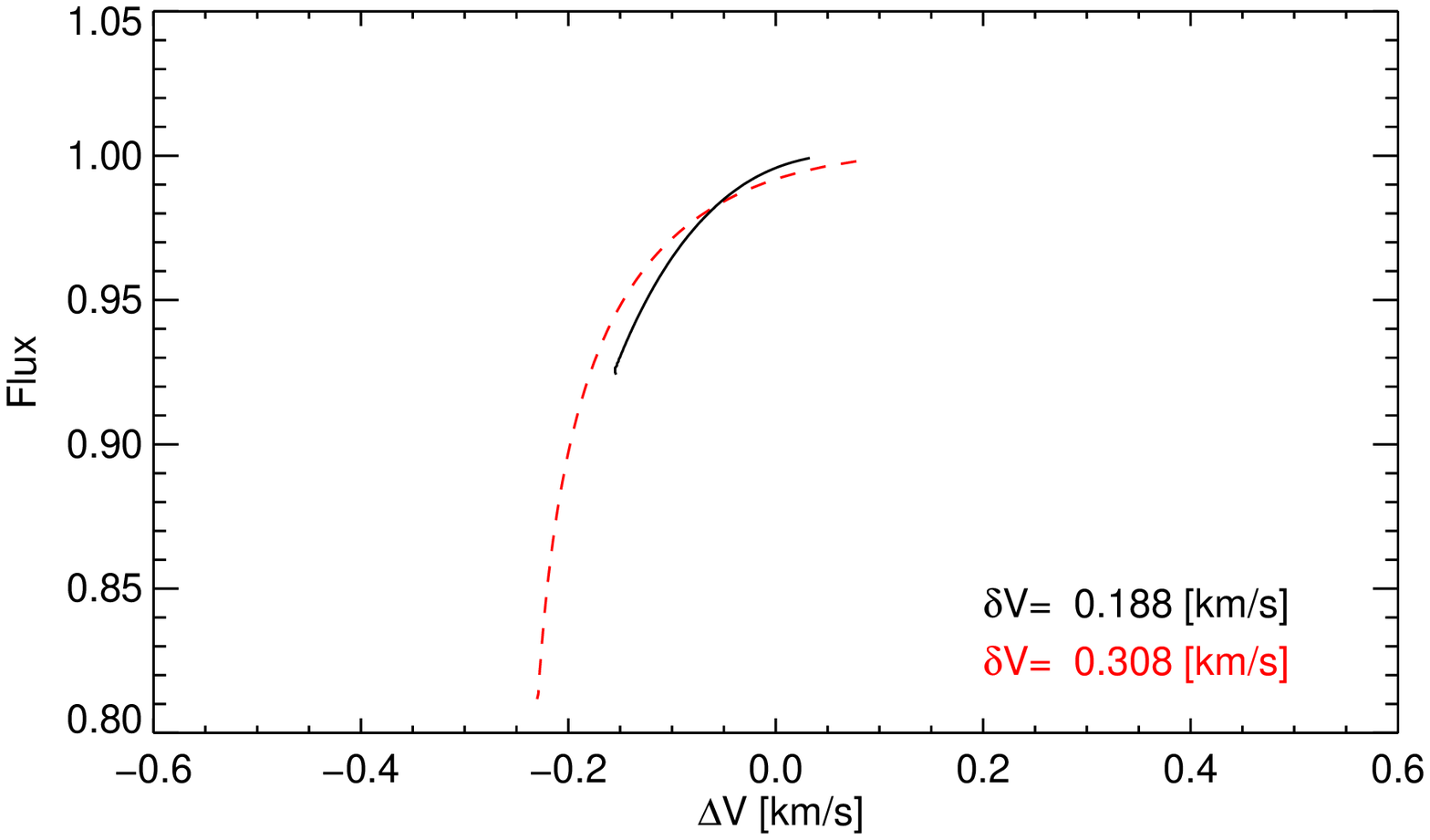}}
\end{center}
% \vspace*{-1.0 cm}
\caption{Comparison of 3D LTE (dashed) and 3D non-LTE (solid) line profile 
(left) and line bisector (right) of a single $^7$Li component, computed for 
a typical metal-poor turn-off halo star ($T_{\rm eff}=6300$~K, $\log g=4.0$, 
[Fe/H] $=-2$). Non-LTE effects strongly reduce the equivalent width of the 
line (W: $35.5 \rightarrow  15.6$~m\AA) while they increase the half width of 
the line profile ($FWHM$: $7.7 \rightarrow 8.5$~km/s); the line asymmetry 
is diminished in non-LTE (velocity span of bisector $\delta v: 0.31 
\rightarrow 0.19$~km/s).}
\label{fig1}
\end{figure}

Finally, 3D non-LTE synthetic line profiles of the Li\,I $\lambda\,6707$~\AA\
feature are computed with the line formation code Linfor3D 
({http://www.aip.de/$\sim$mst/linfor3D\_main.html}), using 
the departure coefficients $b_i = n_i({\rm NLTE})/n_i({\rm LTE})$
provided by NLTE3D for each level $i$ of the lithium model atom as a function 
of geometrical position within the 3D model atmospheres. As demonstrated in 
Fig.\,\ref{fig1}, 3D non-LTE effects are very important for the metal-poor 
dwarfs considered here: not only is the 3D LTE equivalent width too large by 
more than a factor 2, but also is the half-width of the 3D LTE line profile 
too narrow by about 10\%. Moreover, the lithium lines are significantly less 
asymmetric if the non-LTE effects are taken into account.\\[-5mm]

\section{Method and Results}
As outlined above, the $^6$Li abundance is necessarily overestimated if one
ignores the intrinsic asymmetry of the $^7$Li line profile. To quantify this 
error theoretically, we rely only on synthetic spectra. The idea is to 
represent the observation by the synthetic 3D non-LTE line profile of the 
$^7$Li line blend. This 3D flux profile is computed with zero $^6$Li content.
Except for an optional rotational broadening, the only source of non-thermal 
line broadening is the 3D hydrodynamical velocity field, which also
gives rise to a convective blue-shift and an intrinsic line asymmetry.
Next we compute a small grid of 1D LTE synthetic line profiles of the full 
$^6$Li/$^7$Li blend from a so-called 1D LHD model, a 1D mixing-length model 
atmosphere that has the same stellar parameters and uses the same microphysics 
and radiative transfer scheme as the corresponding 3D model. The parameters of
the grid are the total $^6$Li+$^7$Li abundance, $A$(Li), and the $^6$Li/$^7$Li 
isotopic ratio, $q$(Li). Microturbulence is fixed at $\xi_{\rm mic}=1.5$~km/s, 
$v\,\sin i$ is identical to the value used in the 3D spectrum synthesis 
(we tried $0$ and $2$~km/s). Now the 1D line profiles from the grid are 
used to fit the 3D profile. Four parameters are varied independently to find 
the best fit (minimum $\chi^2$): in addition to $A$(Li) and $q$(Li), which 
control line strength and line asymmetry, respectively, we also allow for an 
extra line broadening characterized by $FWHM$ of the Gaussian kernel, and 
a global line shift, 
$\Delta v$. The value $q^\ast$(Li) of the best fit is then identified with 
the correction $\Delta\,q$(Li) that has to be \emph{subtracted} from the 
$^6$Li/$^7$Li isotopic ratio determined from the 1D LTE analysis in order to 
correct for the bias introduced by the intrinsic line asymmetry: 
$\Delta\,q$(Li) =  $q^\ast$(Li), and 
$q^{(3D)}$(Li) = $q^{(1D)}$(Li) - $\Delta\,q$(Li). The procedure
takes saturation effects properly into account.

\begin{figure}[b]
% \vspace*{-2.0 cm}
\begin{center}
\mbox{\includegraphics*[bb=66 38 540 324,width=0.48\textwidth]
{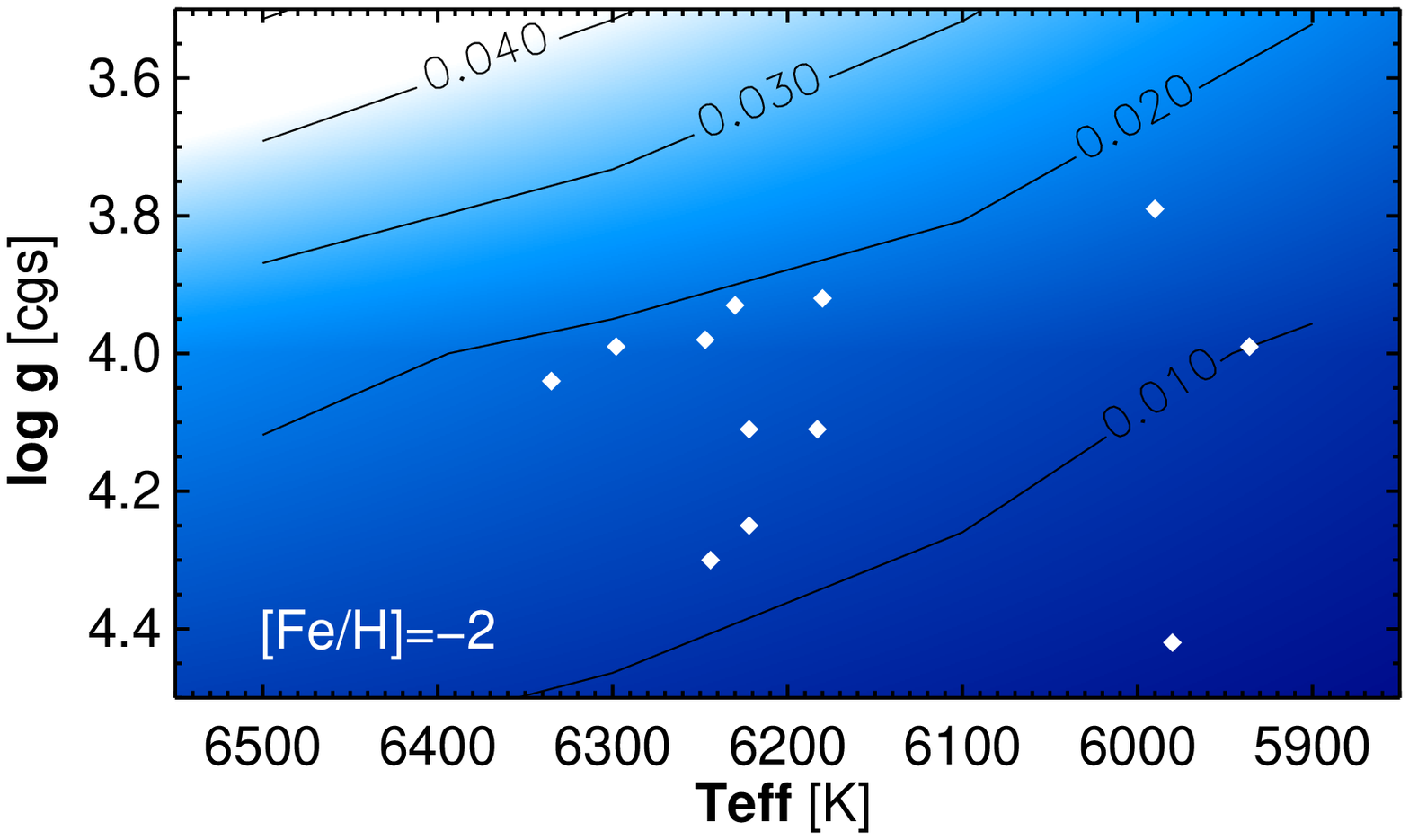}}
\mbox{\includegraphics*[bb=66 38 540 324,width=0.48\textwidth]
{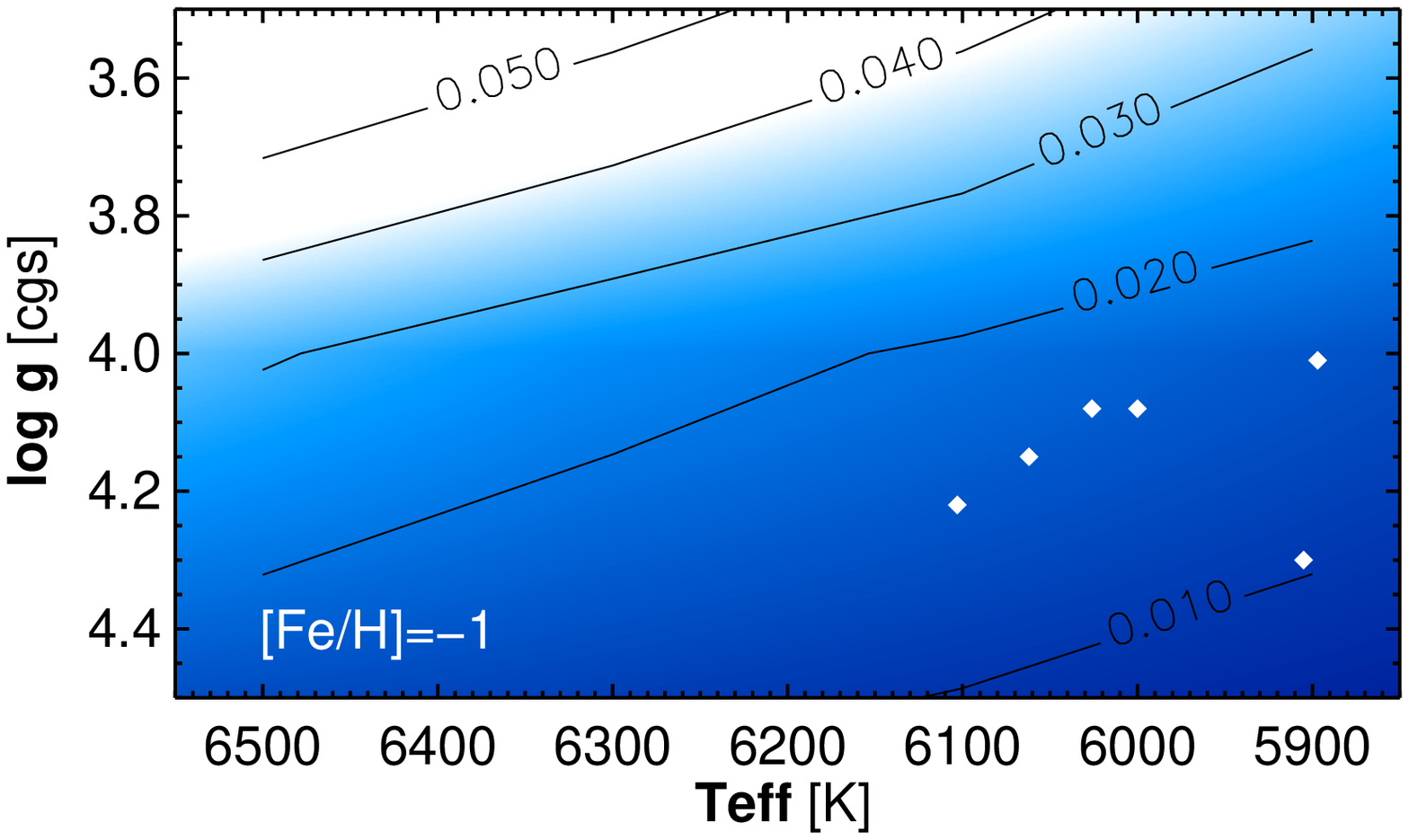}}
\end{center}
% \vspace*{-1.0 cm}
\caption{Contours of $\Delta q$(Li) in the $T_{\rm eff}$ -- $\log g$ plane
for metallicities [Fe/H]=$-2$ (left) and [Fe/H]=$-1$ (right). White
symbols mark the positions of the stars from the \cite{A2006} sample with
$-2.5 <$ [Fe/H] $<-1.5$ (left), and $-1.5 <$ [Fe/H] $<-0.5$ (right).}
\label{fig2}
\end{figure}

\begin{figure}
% \vspace*{-2.0 cm}
\begin{center}
\mbox{\includegraphics*[bb=48 38 560 324,width=0.48\textwidth]
{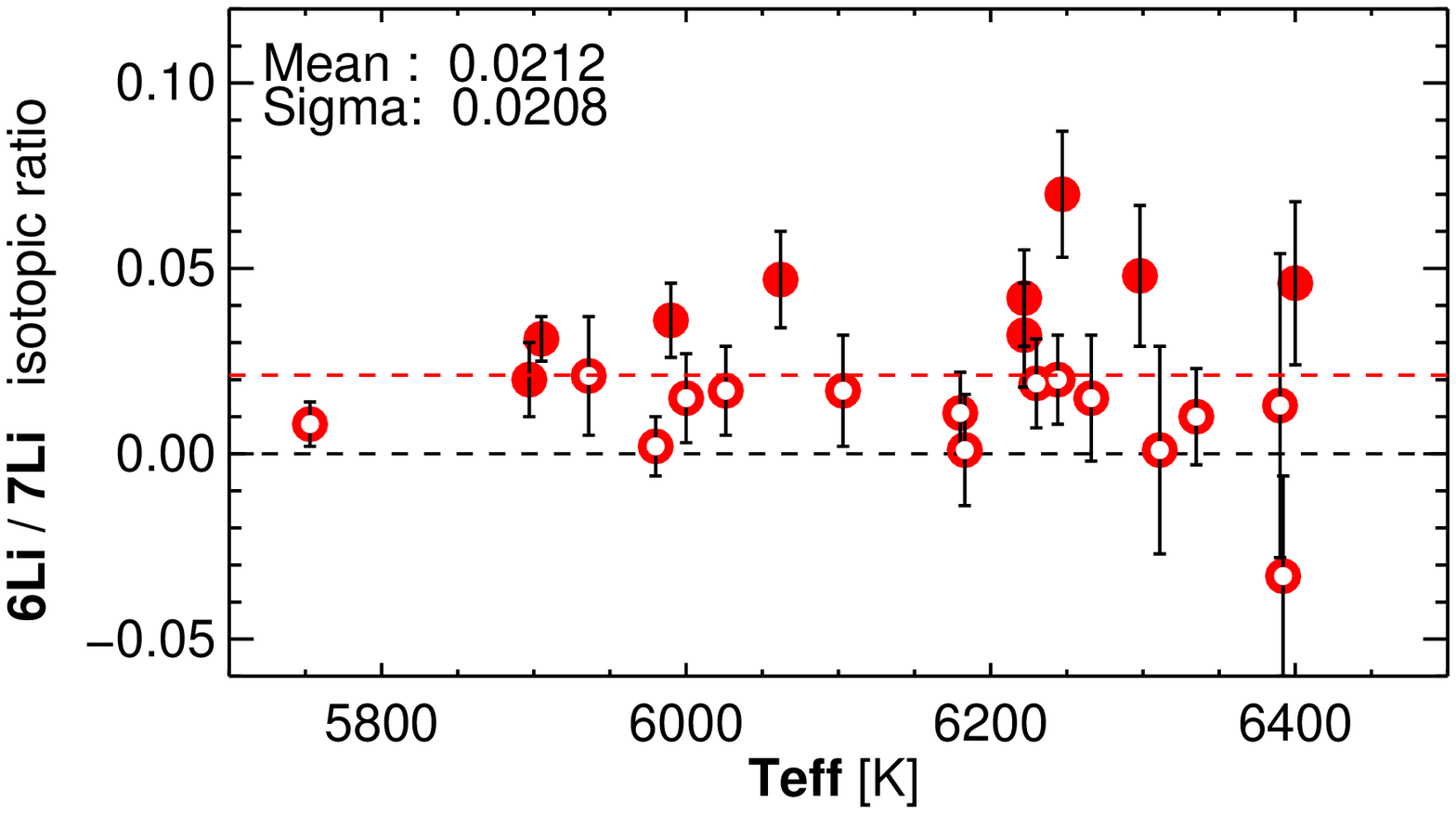}}
\mbox{\includegraphics*[bb=48 38 560 324,width=0.48\textwidth]
{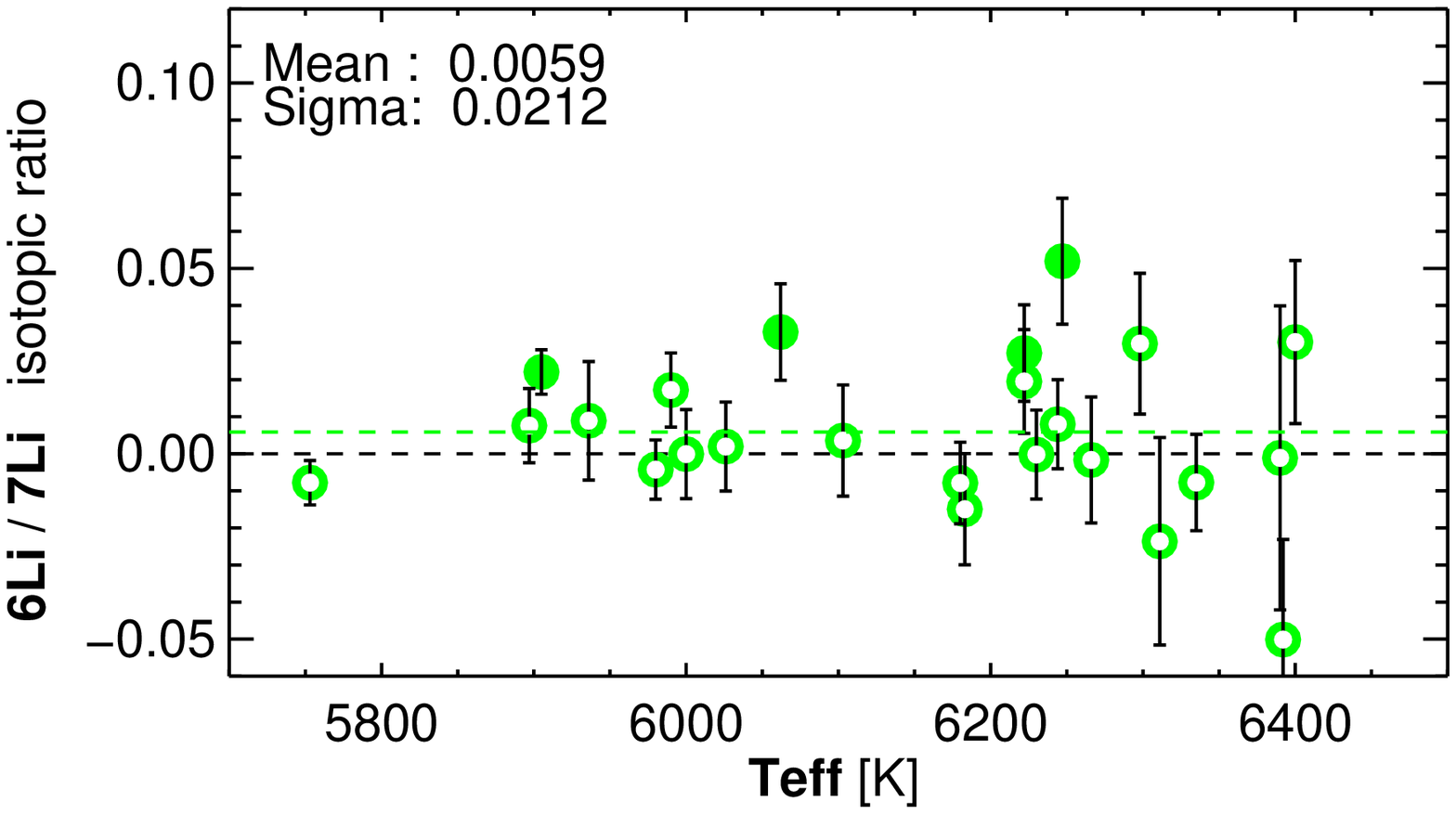}}
\vspace*{-0.1 cm}
\caption{$^6$Li/$^7$Li isotopic ratio, and $\pm 1 \sigma$ error bars, as a 
function of effective temperature as derived by \cite{A2006} before (left) 
and after (right) subtraction of $\Delta q$(Li) to correct for the bias due 
to the intrinsic line asymmetry. 
Filled circles denote $^6$Li detections above the $2\sigma$ level, open 
circles denote non-detections.}
\label{fig3}
\end{center}
\end{figure}

We have determined $\Delta\,q$(Li) in the relevant range of stellar
parameters according to the method outlined above. The results are
displayed in Fig.\,\ref{fig2} for [Fe/H]=$-1$ and $-2$. At given
metallicity, the corrections are largest for low gravity and high 
effective temperature, increasing towards higher metallicity.
We note that $\Delta\,q$(Li) is essentially insensitive to the choice 
of $v\,\sin i$.  The downward correction of the $^6$Li/$^7$Li isotopic
ratio is typically in the range $0.01 < \Delta q$(Li) $< 0.02$ for the
stars of the \cite{A2006} sample (see Fig.\,\ref{fig2}). 
After subtracting for each of these
stars the individual $\Delta q$(Li), according to $T_{\rm eff}$, $\log
g$, and [Fe/H], the mean $^6$Li/$^7$Li isotopic ratio of the sample is
reduced from $0.0212$ to $0.0059$, as illustrated in Fig.\,\ref{fig3}.
If we keep the error bars given by \cite{A2006}, the number of stars
with a $^6$Li detection above the 2$\sigma$ level decreases from $9$ to 
$4$. One of them, HD\,106038, survives only because of its particularly
small error bar of $\pm 0.006$, another one, CD-30~18140, just barely
fulfills the 2$\sigma$ criterion. The remaining two stars are G020-024,
which shows the clearest evidence for the presence of $^6$Li
($q$(Li)~$=0.052 \pm 0.017$), and HD\,102200 with a somewhat weaker
$^6$Li signal ($q$(Li)~$=0.033 \pm 0.013$). The spectra of these stars
should be reanalyzed with 3D non-LTE line profiles.

\begin{table}
  \begin{center}
  \caption{Fitting three observed Li\,I $\lambda\,6707$~\AA\ 
           spectra with 1D LTE and 3D  non-LTE synthetic line profiles,
           respectively. Columns (7)-(10) show the results for 
           $v\,\sin i$ = 0.0 / 2.0 km/s.}
  \label{tab1}
 {\scriptsize
  \begin{tabular}{|l|c|c|c|c|c|c|c|c|c|}\hline
{\bf Star} & {\bf $T_{\rm eff}$} & {\bf $\log g$} & {\bf [Fe/H]} & {\bf S/N} & 
{\bf Model}$^{~1)}$ & {\bf $A$(Li)}$^{~2)}$ & {\bf $q$(Li)} & 
{\bf $FWHM$}$^{~3)}$  & {\bf $\Delta v$} \\ 
{} & {\bf [K]}           &                &              &           & 
{}             &                       & {\bf [\%]}    &  
{\bf [km/s]}         & {\bf [km/s]}     \\ 
\hline
HD\,74000  & $6203$ & $4.03$ & -2.05 & $600$ & 3D NLTE & 
$2.25$ / $2.25$ & {\bf -1.1 / -1.1} & $3.1$ / $2.1$ & ~0.64 / ~0.64 \\ 
 &  &  &  &   & 1D~~~LTE & 
$2.23$ / $2.23$ & {\bf ~0.6 / ~0.6} & $5.9$ / $5.4$ & ~0.42 / ~0.42 \\ 
G271$-$162 & $6230$ & $3.93$ & -2.30 & $550$ & 3D NLTE & 
$2.30$ / $2.30$ & {\bf ~0.6 / ~0.5} & $3.6$ / $2.9$ & ~0.04 / ~0.05 \\ 
 &  &  &  &   & 1D~~~LTE & 
$2.27$ / $2.27$ & {\bf ~2.2 / ~2.2} & $6.2$ / $5.7$ & -0.17 / -0.17 \\ 
HD\,84937  & $6310$ & $4.10$ & -2.40 & $630$ & 3D NLTE & 
$2.21$ / $2.20$ & {\bf ~4.0 / ~4.2} & $3.7$ / $2.7$ & ~0.08 / ~0.07 \\
 &  &  &  &   & 1D~~~LTE & 
$2.18$ / $2.18$ & {\bf ~6.3 / ~6.0} & $6.1$ / $5.6$ & -0.17 / -0.14 \\ 
\hline
  \end{tabular}
  }
 \end{center}
\vspace{1mm}
 \scriptsize{
 {\it Notes:}
  $^{1)}$ $T_{\rm eff}$/$\log g$/[Fe/H] = 6280K/4.0/-2;  $^{2)}$ $\log \left[n(^6{\rm Li})+n(^7{\rm Li})\right] - 
  \log n({\rm H)}+12$; $^{3)}$ Gaussian kernel}
\end{table}

As a consistency check, we have also fitted a few observed Li\,I 
$\lambda\,6707$~\AA\ spectra with 1D~LTE and 3D~non-LTE synthetic 
line profiles, respectively. The fitting parameters are again
$A$(Li), $q$(Li), $FWHM$, and $\Delta v$. As expected, the 3D 
analysis yields lower $q$(Li) by roughly $-0.02$. Details are compiled in 
Table\,\ref{tab1}. HD\,74000 and G271$-$162 are considered non-detections,
while HD\,84937 remains a clear $^6$Li detection with $q^{(3D)}$(Li) 
$\approx0.04$.\\[-8mm]

\section{Conclusions}
The present study indicates that only $2$ or at most $4$ out of the 
$24$ stars of the \cite{A2006} sample remain significant $^6$Li 
detections when subjected to a 3D~non-LTE analysis, suggesting that 
the presence of $^6$Li in the atmospheres of galactic halo stars is 
rather the exception than the rule. This would imply that it is no longer
necessary to look for a global mechanism accounting for a $^6$Li
enrichment of the galactic halo, but that it is sufficient to explain only
a few exceptional cases, which is probably much easier.\\[-5mm]

\end{document}